%% file: ms.tex
\documentclass[sigconf]{acmart}

\usepackage{booktabs} 
\usepackage{hyperref}
\usepackage{graphicx}
\usepackage{mdframed}
\usepackage{subcaption}
\usepackage{soul}

\newcommand{\itund}[1]{\textit{\underline{#1}}}

\newcommand\ra[2]{
\quad
  \begin{mdframed}
    \textbf{Research Answer #1}: \textit{#2}
  \end{mdframed}
}

\newcommand\head[1]{\noindent{\underline{\textbf{#1 }}}}
\newcommand\OSS[1]{$\mathcal{OSS_{#1}}$}
\newcommand\ISS[1]{$\mathcal{ISS_{#1}}$}
\newcommand\DAF{$\mathcal{DAF}$}

\newcommand\rqone{\noindent\textbf{RQ 1: Can we describe/predict future @-mentions in terms  of developer visibility, expertise, productivity, and responsiveness? }}
\newcommand\rqtwo{\noindent\textbf{RQ 2: Can models trained entirely on one project be reliably used to predict @-mentions on another project? }}
\newcommand\rqthree{\noindent\textbf{RQ 3: Is there evidence of project-specific @-mention culture? Or are the affecters of @-mentions a GitHub-wide phenomenon? }}

\newcommand\etal{\emph{et al. }}
\newcommand{\etc}{\emph{etc. }}

\begin{document}
\title{Whom Are You Going to Call?: Determinants of\\ @-Mentions in GitHub Discussions}

\author{David Kavaler}
\affiliation{%
  \institution{University of California, Davis}
  \city{Davis} 
  \state{California} 
}
\email{dmkavaler@ucdavis.edu}

\author{Premkumar Devanbu}
\affiliation{%
  \institution{University of California, Davis}
  \city{Davis} 
  \state{California} 
}
\email{ptdevanbu@ucdavis.edu}

\author{Vladimir Filkov}
\affiliation{%
  \institution{University of California, Davis}
  \city{Davis} 
  \state{California} 
}
\email{vfilkov@ucdavis.edu}

\begin{abstract}
\input{abstract}
\end{abstract}

%
%
%

\maketitle

\section{Introduction}

In modern, social-coding~\cite{dabbish2012social} projects based on sites like GitHub and BitBucket, that favor
the pull-request model, the emergence and growth of a particular socio-technical link type, @-mentions, can be observed in task-oriented technical discussions.
For example, in the \emph{rails} project on GitHub (issue $31804$), one of the head developers calls on another,
explicitly stating trust of their expertise, saying: \emph{``@kamipo can you take a look since you are our MySQL expert?''}
On GitHub, the @-mention in issue discussions is a type of directed social link; the
@-mentioner causes a directed communication to be sent to the @-mentionee
through GitHub's interface. Thus,
one can consider the network of @-mentions, specifically \emph{calls}, as a sort of a directed social network, with a task-oriented purpose.
These mentions are heavily used in social coding; in our data,
$52.46\%$ of issues and $22.02\%$ of pull requests contain at least one @-mention, with an average of $1.46$ and $1.37$ @-mentions per issue or pull 
request (respectively). On average, developers who are called (while not yet actively participating in the thread) respond $19\%$ of the time;
the number rises to   \emph{42.94\%} when
excluding those who never respond\footnote{\emph{E.g.}, developers of upstream libraries rarely respond in the downstream project.}.
@-mention ubiquity reflects the central role they play in task-oriented social interactions. 
Since much of a developers behavior in OSS projects is recorded,  
if a person has the expertise and/or are reliable in many different tasks, they will be visible to others. 
The decision to @-mention someone will be based on visible attributes of that developer, including reliability, productivity, \etc 
Identifying a reliable and knowledgable person to ask for help or action is key to addressing issues in a timely manner and keeping a project vibrant and alive. In fact, Yu \etal found that having @-mentions in a discussion decreases the time to resolve an issue~\cite{yu2016reviewer};
Zhang \etal found that more more difficult issues (\emph{e.g.}, longer length of discussion) have more @-mentions~\cite{zhang2017social}.
Given these important outcomes, it would be beneficial to know what (observable) socio-technical attributes of developers contribute 
to being @-mentioned.


As @-mentions have an inherent social element, a global model describing the affecters of
@-mention calls would suggest that project-specific social idiocyncracies are less important than social elements common across GitHub. A global
GitHub model for @-mentions may be seen as positive, as shared social norms across large populations can increase social 
mobility~\cite{sato2004impact}; on GitHub, this may make the acculturation process easier for those who move between projects. In addition,
the findings of Burke \etal~\cite{burke2009feed} suggest that those who perceive themselves as socially central contribute more as a result - this
may extend to code contributions on GitHub. The
findings of Kavaler \etal~\cite{kavaler2017perceived} suggest global and project-specific social phenomena (apropos language use) exist
on GitHub; is this the case for @-mentions? Or does one phenomenon dominate?


The goal of this paper is to understand both the elements contributing to @-mentions in GitHub projects and the extent to which those elements are shared between projects across GitHub.
@-mentioning is a complex, multidimensional phenomenon.
Developers that are often @-mentioned can have outsized roles and responsibilities in the project network, and be able to handle any task.
Thus, @-mentioning someone means, first of all that they observably contribute and have contributed to the project, but also that they can be relied on, or maybe even trusted, with the task at hand. 
Whereas visibility can be operationalized more directly, based on a person's aggregate presence in all aspects of the social coding process, both reliability and trust are more complex; we describe the theoretical background for these in the next section.
Starting from those theories, and from data  on @-mentions and comprehensive developer and project metrics from 200 GitHub projects, we sought to quantitatively model future @-mentions of a developer predictively, from past observations of the developer's visibility, expertise, productivity, and responsiveness in their projects.
From our quantitative models, together with case studies aimed towards triangulating the model results, 

\begin{itemize}
\item 
We find that we can mine a reliable @-mention signal from GitHub data as per existing theoretical definitions in sociology, psychology, and management.

\item We see a net positive effect of visibility on @-mentions. We see that less expertise (via, \emph{e.g.}, commits that need fixing, likely buggy commits) associates with lower
@-mentions when one has already been @-mentioned, and higher @-mentions if one has not already been @-mentioned; perhaps explained 
by the idea that any productivity, even buggy, associates with an initial @-mention, consequently adjusted. 
We see positive effects for productivity, and none for responsiveness.

\item We find that cross-project model fits are generally good, suggesting a common model of @-mentions across GitHub.
Similarities among the models are greater for enhanced @-mentions after the first @-mention, than for the initial one.

\item We see indications of project-specific @-mentioning behavior, however, the high performance of cross-project prediction
suggests the differences may matter little, especially for predicting @-mentions.

\end{itemize}

We present the theory and research questions in Sect.~\ref{sec:theory},
data and methods in Sect.~\ref{sec:data_methodology},
results and  discussion in Sect.~\ref{sec:results_and_discussion}, and the threats and conclusion in Sect.~\ref{sec:conclusion}.

\section{Theory and Related Work}
\label{sec:theory}

To understand the notion of @-mentions in OSS projects, we build a theory drawing from diverse sources.
First, we discuss  @-mentions and their use on GitHub, supported by prior work. Then, we introduce theory behind GitHub @-mentioning drawn from work regarding reliability and trust in the fields of sociology, psychology, and management. We then discuss the importance of social exchange and interaction (and thus, the importance of @-mentions) on OSS project success.

\head{@-Mentions on GitHub}
GitHub projects have issue trackers with a rich feature set, including ticket labeling, milestone tracking, and code tagging.
In GitHub projects, individuals can open up an issue thread where others can comment and discuss a specific issue.
In these discussions, developers can tag others using \emph{@-mentions}; the mentioned developer  receives a notification that they are being referenced in a discussion.
When one decides to @-mention another developer, there is generally a specific reason, \emph{e.g.},
to \emph{reply} to a single person in a discussion involving many others; or, to \emph{call} the attention of someone who isn't currently in the discussion.
The latter aspect is what we wish to capture; calling upon another person is an implicit (and on GitHub, often explicit) statement of belief
that the receiver
will be useful in addressing the task at hand.
To validate the importance of modeling \emph{call} @-mentions on GitHub, we perform a case study (Section~\ref{sec:case_study}) 
and also look to prior literature (below) for the reasons behind the use of @-mention.

Tsay \etal performed interviews with several developers of popular projects on GitHub, specifically related to the discussion and
evaluation of contributions~\cite{tsay2014let}. They found that both general submitters and core members use @-mentions to alert
core developers to evaluate a given contribution or start the code review process. They further found that core members
often @-mentioned other core members specifically citing that the @-mentionee is more qualified to answer a particular question or review
a given contribution. In nearly all cases, the @-mention seems to be used to draw the attention of a developer who may contribute
to the task at hand.
Kalliamvakou \etal surveyed and interviewed developers, mostly commercial, that use GitHub for development~\cite{kalliamvakou2015open}.
Of all interviewees, $54\%$ stated
that their first line of communication is through the @-mention\footnote{Developers were asked about communication methods, not explicitly the @-mention.}.
In addition, they state that teams often use the @-mention to draw
members' attention to a problem.
\footnote{Described in Sect.~\ref{sec:data_issues_trust}, a reply @-mention
is directed towards someone already in the discussion; a call @-mention is directed towards someone not yet in the discussion.
In our data, there is indeed a very high correlation between reply @-mentions and discussion
length ($81.16\%$); however, there is a relatively low correlation between call @-mentions and discussion length ($28.29\%$). As our
focus is on call @-mentions, correlation between reply @-mentions and discussion length is not a threat.}


\head{@-Mentions and Personal Reliability}
The ability to rely on others socio-technically is critical for cohesive workgroups. 
From a social perspective, Saavedra \etal argue
that reliable interactions among group members are important for success, especially when tasks are interdependent~\cite{saavedra1993complex}. 
According to social learning theory, frequent interactions among group members increases the likelihood that some in the group will be raised to
``role model'' status~\cite{bandura1973aggression,bandura1977social}. The importance of role models in social learning has been widely
discussed~\cite{bandura1977social,burke2009feed,dourish1994running}. On GitHub, researchers have found that these role models (``rockstars'')
are important influencers, allowing developers to learn from ``rockstar'' code contributions in order to improve 
their own work~\cite{dabbish2012social,lee2013github}.
In other words, developers rely on others within and outside  their immediate working group in order to solve problems. 
In addition, peer code review (relying on team members other than the authors for manual inspection of source code) is recognized as a valuable tool in
software projects~\cite{ackerman1984software,ackerman1984software}.
Thus, we argue that identifying
these reliable developers, by means of the @-mention, is important for project success.
We theorize that reliability will manifest itself on GitHub through
\itund{responsiveness}, measuring: if you are called, how often do you answer?

\head{@-Mentions and Trust}
Trust has a long-recognized complex~\cite{mcknight2002developing,gallivan2001striking} social component and well understood benefits to social and economical well-being~\cite{inglehan1999trust, newton2001trust}, in both physical and virtual teams~\cite{jarvenpaa1998anybody}.
While individuals do have a personal notion of when to trust someone, in social settings those notions inherit from the communal sense of trust~\cite{newton2001trust, jarvenpaa1998anybody, inglehan1999trust}.
In socio-technical groups like software projects, contributors must be trusted as technically competent, and also as useful to the project.
Gaining contributor status is a key indicator of trust worth careful study; considerable prior work has done so~\cite{bird2007open, steinmacher2015social, casalnuovo2015developer,gharehyazie2013social,ducheneaut2005socialization}.
In pull-request oriented models, with \emph{decentralized} repositories, code contributions can be made to forks,
then packaged as a pull-request, without restraint.
Here, social processes such as code-review take a central role in deciding the fate of code contributions.  
Opinions from trusted people during the relevant discussions would  be in great demand, and thus, the social demand on a person is an indication of the trust placed upon them by the community.
Since the pull-request model is more or less normative in GitHub projects, 
it is reasonable to posit that many projects in the GitHub community ecosystem may share the same determinants @-mention extension, \emph{i.e.}, the reasons behind @-mention extension may be a global phenomenon. 

We acknowledge the extension of an @-mention is not necessarily due purely to trust in the taggee; however,
some form of trust likely plays a role. Thus, understanding theories of trust is important to understanding
@-mentions on GitHub.

Oft-mentioned and widely discussed, the meaning and role of trust has been examined across many disciplines, including
sociology, psychology, and philosophy~\cite{zucker1986production,brockner1996understanding,kramer1996trust,brenkert1998trust,husted1998ethical}. 
Gallivan provides a succinct set of definitions for trust types as provided by prior work on organizational
trust~\cite{gallivan2001striking}; relevant types for GitHub are: 1) \emph{Knowledge-based trust}: trust based upon a prior history of transactions between two parties; 2) \emph{Characteristic-based trust}: trust that is assumed, based on certain attributes of the other party; and 3) \emph{Swift trust}: a ``fragile'' form of trust that emerges quickly in virtual workgroups and teams.


For our work, the idea of swift trust is important as it is theoretically defined for virtual teams, as
on GitHub. Jones and Bowie~\cite{jones1998moral} state: ``\emph{the efficiency of [virtual teams] depends on features - speed and flexibility -
that require high levels of mutual trust and cooperation}''; other researchers share and expand on this notion~\cite{o2002distributed,handy1995trust}.
Though swift trust may initially appear most applicable, much of the founding work was
done prior to the proliferation of socio-technical 
systems such as GitHub.
More recently, Robert \etal redefine swift trust for modern systems as a combination of classical swift trust, knowledge-based trust, and parts of characteristic-based trust~\cite{robert2009individual}.
We agree with this blended definition - a sweeping categorization of GitHub as having a swift trust system is likely incomplete; multiple trust
regimes probably apply. 
We capture knowledge-based trust through our measures of
\itund{visibility}, \emph{i.e.}, functions of @-mention network degree.
Characteristic-based trust is also likely; task characteristics can be easily
seen on GitHub, as captured by measures of \itund{expertise} and
\itund{productivity}.




\head{@-Mentions and Social Exchange}
On GitHub, the @-mention is a type of directed social link; the @-mentioner causes a notification to be sent to the @-mentionee
through GitHub's interface, a form of social communication. Thus, the network of @-mentions is a sort of social network, with a
task-oriented purpose.
Much work has been done in variety of fields on identifying reasons behind social tagging and mentioning behavior, including
on GitHub~\cite{yu2014exploring}.

In the fields of psychology and sociology, many researchers have explored the phenomenon of social tagging on
Facebook~\cite{qiu2013cultural,burke2010social,oeldorf2015posting}. In general, this research has shown
that social tagging provides a sense of community and increases one's social capital.
These findings are of importance to GitHub as they elucidate
the importance of community social interaction,
which are known to be important to OSS success~\cite{gharehyazie2015developer,gharehyazie2013social}.
Of specific interest, Burke \etal found that those who receive feedback on their Facebook
posts share more~\cite{burke2009feed}. It is reasonable to believe that this extends to task-oriented networks, such as GitHub;
those who feel as though their contributions are important, socially or technically, are likely to contribute more.

McDonald \etal interviewed multiple GitHub developers
and found that they rarely use product-related measures (\emph{e.g.}, release quality, bug fixes) to describe project success; rather, they use
measures such as number of (new) contributors, pull requests, \emph{etc}~\cite{mcdonald2013performance}.
As stated above, social exchange is important to both one's own well-being and OSS success. As social measures have been shown to be important
for OSS \emph{product} success~\cite{hossain2009social}, and given that developers generally use non-product measures
to describe \emph{project} success, fostering the use of @-mentions and thus the exchange and gain of social capital would be beneficial for
both metrics of success. We capture social aspects in \itund{visibility} - functions of @-mention network measures.

\head{@-Mentions and Discourse/Dialogue}

Discourse and dialog have seen a resurgence of research interest with the advent of NLP computational methods.
 Stolcke \etal~\cite{stolcke2000dialogue} have most prominently defined discrete
conversational speech categories into which 
@-mentions fit well, perhaps because they themselves are social link extensions.
Stolcke's \etal~\cite{stolcke2000dialogue} work and the other aforementioned prior work~\cite{tsay2014let,kalliamvakou2015open},
helped us distill the following four categories of speech that use @-mentions (one of these is a slightly modified category as compared to Stolcke's work, marked by $\star$):

\begin{enumerate}
\item \textbf{Request (R)}: An explicit request towards the called person to perform some action.

\item \textbf{Request-Suggest (R-S)} : An implicit request towards the called person to perform some action.

\item \textbf{Inform (I)}: An indication that the issue or post is relevant to the called person.

\item \textbf{$\star$Credit Attribution (CA)}: An @-mention designed to attribute credit to the called person. This is similar to ``Thank'' by Stolcke \etal~\cite{stolcke2000dialogue}, but explicitly directed at an individual.
\end{enumerate}

We use these categories in a case study examining reasons behind call @-mentions in Section~\ref{sec:case_study}.

\subsection{Research Questions}
\label{sec:research_questions}
@-mentions 
signal a desire for a developer's involvement in a task-oriented discussion.
GitHub is a rich source of mine-able, potentially relevant, developer characteristics. 

The theory above allowed us to identify relevand dimensions along which to model the phenomenon of @-mentions.
We describe them shortly here, and operationalize them in the Methods section. \emph{Visibility} 
measures the ability of others to know of a developer; 
if a developer is to be @-mentioned, people must know the network 
in order to know who they are capable of reaching. 
\emph{Expertise} can be defined through task-related measures, \emph{e.g.}, number of likely buggy commits,
which might  influence how much a developer is @-mentioned. 
\emph{Productivity} is defined by number of commits; prolific committers could be viewed as the ``top brass'' of a project, and commits
are easy to see in GitHub. Finally, we are interested in \emph{responsiveness}; if a mentionee is called to lend their talent, it is not farfetched
 that those who respond to the call are more likely to be @-mentioned in the future.

We explicitly model \emph{future} @-mentions, \emph{i.e.}, @-mentions as measured $6$ months beyond the ``observation period'', described further
in Section~\ref{sec:data_mention_model}. Having an effective model that explicitly predicts future behavior has higher utility
to potential future applications than an aggregate regression model over the whole history.

\rqone

Our second question relates to the utility of our model. If one wishes to use our model on their own projects, it would be helpful to be able
to use the model pre-trained on some data, \emph{e.g.}, trained entirely on a separate project and applied to one's own.

\rqtwo

Our third question is more theoretical in nature. Specifically, we wish to describe the differences between projects in terms of our
affecters of @-mentions and identify some potential reasons behind these differences. As GitHub is composed of subcommunities which may
have some idiosyncracies, we believe that these differences may be reflected in our describers of @-mentions.

\rqthree
 

\section{Data and Methodology}
\label{sec:data_methodology}
All data was collected by querying GitHub's public API using the Python package PyGithub\footnote{\url{https://github.com/PyGithub/PyGithub}},
with the exception of issue fixing data, which was gathered by cloning individual repositories (see below).

\head{Filtering and Cleaning}
Our data set started as a sample of $200$ projects from the top $900$ most starred and followed projects. The number of stars and followers are proxies for project popularity, and can identify projects likely to contain enough issues and commits to model robustly. As some measures are expensive to calculate, and we wanted a mixture of high-popularity and medium-popularity projects, we decided to start with a $200$ project sample to avoid skew towards the upper or lower ends of popularity within the $900$.

We ran multiple parallel crawlers on these $200$ projects to gather commits, issues, pull requests, and associated metadata.
Due to some internal issues with the PyGithub package\footnote{PyGithub
did not handle properly some Null responses from GitHub's API.}, some projects failed to
return the entirety of the data. We created a verification system (completely external to PyGithub) to determine which
projects were incomplete, and removed them from consideration. Finally, we only consider developers with at least one commit to their given project
in order to avoid a proliferation of zeros in our covariates, as many developers participate in issue discussions but never contribute.
This was done in order to focus on those who may become @-mentioned in the future; without any commits, we argue it is unlikely
to be @-mentioned in the future.

As we wish to explicitly model future @-mentions, we introduce a time split in our data. For each project, we define a time frame
under which we ``observe'' the project and its participants, and model our response as calculated beyond our observation
time frame - the ``response'' period.
We decided to set our response period as $6$ months, \emph{i.e.}, $6$ months prior and up to the end of our data. We also tested periods of
$3$ and $12$ months; $3$ months had little difference to $6$ months, and $12$ months left us with too little data to model. We then filtered out
each individual who had a project participation shorter than $3$ months in order to have confidence that their
participation has had a chance to stabilize. Thus, we explicitly model future @-mentions, as our response period is 
disjoint from our observation period.
In total, this yielded $154$ unique projects comprised of $17,171$ project-developer pairs to test our hypotheses.

\begin{figure}
\includegraphics[width=0.6\columnwidth]{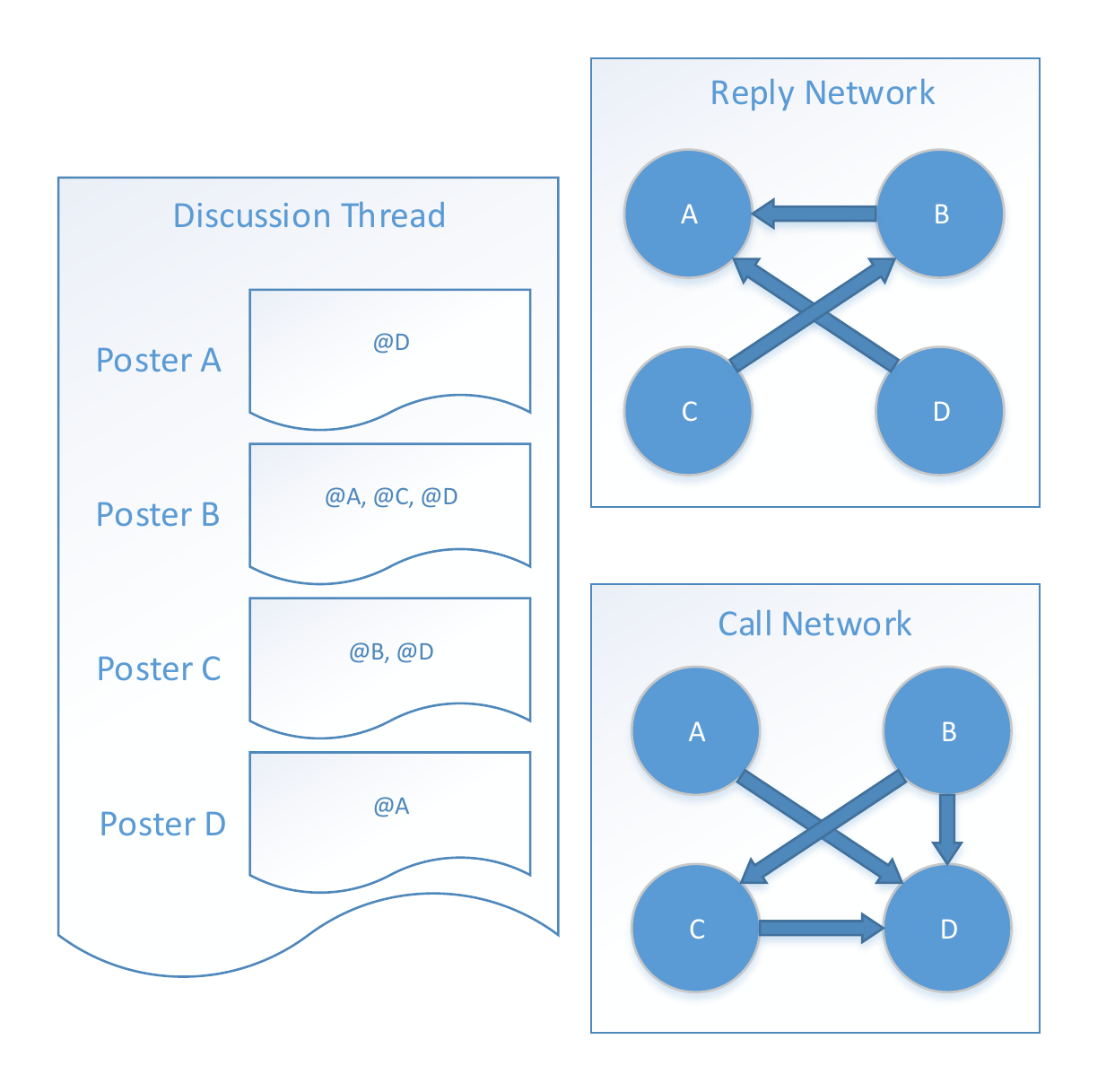}
\caption{The network creation process. Shown is a discussion thread and the resulting reply and call networks. Note this can be
a multigraph (not shown).}
\label{fig:network_creation}
\end{figure}

\head{Issues and @-Mentions}
\label{sec:data_issues_trust}
For each project on GitHub, individuals can open up an issue thread where others can comment and discuss a specific issue\footnote{Note that pull 
requests are a subset of issues.}.
We constructed a social network for each project using @-mentions in their issue comment threads; Fig.~\ref{fig:network_creation}
depicts this process.
Similar to Zhang \emph{et al.}~\cite{zhang2015exploring}, \emph{i.e.}, every edge ($u, v$) is developer $u$ @-mentioning $v$ somewhere in
their post. This yields a directed multigraph; there can be multiple edges ($u, v$) depending on how many times $u$ @-mentions $v$.
We distinguish between two edge types: \emph{reply} and \emph{call}. A \emph{reply} edge is defined by $u$ @-mentioning
$v$ when $v$ \emph{has already posted in the given thread}.
A \emph{call} edge is defined by $u$ @-mentioning $v$ when $v$ \emph{has not yet posted in the given thread}. Thus, a call edge
is representative of the phenomena we wish to measure, described in Sect.~\ref{sec:theory};
$u$ calls upon $v$ as $u$ wishes for $v$'s input for the discussion at hand.

\head{Focus}
As a measure of visibility, we wished to capture phenomena more nuanced than merely raw indegree and outdegree\footnote{Though we do use outdegree in our model as well.}, as raw degree counts do not take into consideration the larger, neighborhood view. Standard global measures used in social
network analysis are often too expensive to calculate for our large @-mention networks. Thus, we require a measure that takes into account
a more global view that is relatively inexpensive to calculate. Here, we introduce the idea of \emph{social focus} in the @-mention network.

Theoretically, we believe that when given many choices on who should be contacted (@-mention), individuals must make a  decision, based on
their knowledge of the potential receiver's characteristics (\emph{e.g.}, ability to help in a task) and who is more readily visible.
In social networks, knowledge of others is propagated through existing links. Thus, if an individual is highly focused-on, it is likely
that they will become more so in the future. This means that the more focused-on a developer is, the more visible they likely are. In addition, those who
have lower social focus on others, \emph{i.e.}, they distribute their out-links widely among many others, are also more likely to be visible to others.


To represent focus, we adapt a metric described by Posnett \emph{et al.}~\cite{posnett2013dual}.
This metric is based on work by theoretical ecologists, who have long used Shannon's entropy to measure diversity - and its dual, specialization - within a species~\cite{good1953population}, and can be derived from Kullback-Leibler divergence. For discrete probability distributions $P$ and $Q$, Kullback-Leibler divergence ($KL$) is defined as:

\vspace{-1.5ex}
\begin{equation*}
D_{KL}(P | Q) = \sum_i {P_i \ln{\frac{P_i}{Q_i}}}
\end{equation*}

Bluthgen \emph{et al.} define a species diversity measure, $\delta$\footnote{This measure is originally called $d$ by Bluthgen \emph{et al.},
but we will use $\delta$ here to reserve $d$ to represent developers.}, using $D_{KL}$~\cite{bluthgen2006measuring}.
This measure is calculated naturally in a bipartite graph formulation, where each species in the graph has its own diversity
value $\delta_i$. Posnett \emph{et al.} use this metric, normalized by the theoretical maximum and minimum (\emph{i.e.},
so $\delta_i$ ranges from $0$ to $1$), to measure ``developer
attention focus'' ($DAF$)~\cite{posnett2013dual}. 
%
When $\delta_i$ (a row-wise measure) is high, developer $i$ is more focused in commits to a fewer number of modules.
Analogously, when $\delta_j$ (a column-wise measure) is high, module $j$ receives more focused attention from fewer developers.
%
They call these quantities ``developer attention focus'' ($\mathcal{DAF}_i$) and ``module attention focus'' ($\mathcal{MAF}_j$)\footnote{We do not use $\mathcal{MAF}$, we use an analogous form for our social networks.}.

In this work, we take these definitions and expand them to the social network of @-mentions. Recall that we distinguish between
two types of @-mentions: \emph{reply} and \emph{call}. We can likewise represent our social network as a bipartite graph, where
the rows and columns of the adjacency matrix both refer to developers, and each cell $s_{uv}$ is the count of directed @-mentions from developer $u$
to developer $v$ for a given @-mention type. Thus, we analogously define $\rho_u$ as the focus developer $u$ gives in their reply @-mentions,
and $\rho_v$ as the focus developer $v$ receives from others' reply @-mentions.
Similarly, we define $\kappa_u$ as developer $u$'s focus in their call @-mentions,
and $\kappa_v$ as the focus developer $v$ receives from others' call @-mentions.

Recall that we can interpret these values equivalently as a measure of \emph{specialization} or \emph{inverse uniformity}. For example, if $\rho_u$ is large,
developer $u$ specializes their replies to a select group of others; if $\rho_u$ is small, developer $u$ uniformly replies to all others. Likewise, if
$\kappa_v$ is large, developer $v$ is called by a select group of others; if $\kappa_v$ is small, developer $v$ is called uniformly by all others.
We believe this intuition
is useful to answer our research questions.
Thus, we define normalized \emph{outward social specialization} and \emph{inward social specialization} measures for both replies ($\rho$) and calls ($\kappa$):

\vspace{-2ex}
\begin{equation*}
\begin{aligned}[t]
\mathcal{OSS}_{u,\rho} = \frac{\rho_u - \rho_{u,min}}{\rho_{u,max} - \rho_{u,min}}
\end{aligned}
\qquad
\begin{aligned}[t]
\mathcal{ISS}_{v,\rho} = \frac{\rho_v - \rho_{v,min}}{\rho_{v,max} - \rho_{v,min}}
\end{aligned}
\end{equation*}

\noindent where $\mathcal{OSS}_{u,\kappa}$ and $\mathcal{ISS}_{v,\kappa}$ are defined analogously.

\head{Attributing Commits That Need Changing (likely buggy)}
To identify commits that had to be changed in order to close an issue (\emph{i.e.}, likely buggy commits), we use the standard SZZ algorithm~\cite{sliwerski2005changes}, as expanded in~\cite{kim2006automatic},
with a few changes to accommodate GitHub nuances. GitHub has a built-in issue tracking system;
developers close open issues by using a set of
keywords\footnote{\url{https://help.github.com/articles/closing-issues-using-keywords/}} in either the body of their pull request or commit message.
\emph{E.g.}, if a developer creates a fix which addresses issue $\#123$, they can submit a pull request containing the phrase ``closes $\#123$'';
when the corresponding fixing patch is merged into the repository, issue $\#123$ is closed automatically. To identify likely bug-fixing commits,
we search for associated issue-closing keywords in all pull requests and commits. We then ``git blame'' the respective fixing lines
to identify the last commit(s) that changed the fixing lines, \emph{i.e}, the likely buggy lines. We assume the latest change to the fixing
lines were those that induced the issue, and refer to those changes as likely buggy, or buggy for short.

We note that an issue is a rather broad definition of a bug, as an issue can be brought up to, \emph{e.g.}, change the color of text
in a system's GUI; this may not be considered a bug by some definitions.
However, as GitHub has the aforementioned automatic closing system, we believe that our identification of fixing commits (and therefore buggy commits)
does not contain many false positives.
Prior work has relied on commit message keyword search, which may introduce false positives due to project-level
differences in commit message standards, \emph{i.e.}, what a commit message is expected to convey. These standards
can vary widely~\cite{bird2009promises}.

\head{Variables of Interest}
We are interested in measuring and predicting @-mentions as a function of \emph{readily observable} developer attributes, namely \emph{visibility},
\emph{expertise}, \emph{productivity}, and \emph{responsiveness}. We operationalize these attributes as follows:

We define {\textbf{visibility}} as the ability for developers to note a person's existence; if developer $A$ is not aware of the existence
of developer $B$, it is unlikely that $A$ would @-mention $B$. This is akin knowledge-based trust.
Here, we use our social specialization measures \underline{\OSS{\rho}},
\underline{\OSS{\kappa}},
and \underline{\ISS{\kappa}}, along with \ul{total social outdegree}
(total number of @-mentions for a developer in a given project) as measures of visibility. We believe
these measures are reasonable as they identify one's existence within the social network.

We define \textbf{expertise} as a developer's ability to complete project tasks in accordance with team expectations, related to
characteristic-based trust. 
To represent this, we use \ul{number of issue-inducing commits} made by a developer, \ul{focus measure \DAF},
and a factor identifying whether or not the given developer
is the \ul{top committer or project owner}. A higher number of issues can indicate a lack of aptitude for programming according 
to the project's goals\footnote{We use issues fixed before closing as proxy for bugs; a higher value need not imply lack of aptitude, but 
it indicates a change in expected coding behavior and expertise.}.
It has been shown that a higher \DAF (\emph{i.e.}, higher module specialization) is associated with
fewer bugs in a developer's code~\cite{posnett2013dual}. Thus, \DAF can represent developers' expertise in code modules. The top
committer or project owner factor indicates a certain level of prestige and expertise; one would expect the top contributor 
or project owner would likely be the most expert in matters concerning the project. Number of fixing commits was also calculated,
but was not used due to collinearity with of bug commits.

We measure \textbf{productivity} as the \ul{raw commit (authoring) count}.
Measures of productivity abound-- most
have been shown (of those we computed, \emph{e.g.}, lines of code added or deleted) to highly correlate with commit count, especially in models where confounds are recognized. We choose commit count as it is the simplest.

We describe \textbf{responsiveness}  as a measure to answer the question: when you are called, do you show up? One would expect
that those who are responsive, and thus display their reliability,
will be called upon again.
This is precisely defined as the \ul{number of times a developer is called and responds to that call}; \emph{e.g.}, if a developer is called
in $10$ unique issues and responds in $8$ of those issues, their responsiveness value is $8$.

\textbf{Extra-Project Controls}:
As stated, our interest is to identify \emph{readily observable} attributes of potential @-mentionees (\emph{e.g.},
within-project social activity and commit activity), and functions thereof.
This is in contrast to things that may be hard to observe, such as 
activity outside the project at hand (\emph{e.g.}, outside-project social activity, exact number of commits to other projects, \emph{etc.}).
However, such a control for outside experience is likely necessary as, \emph{e.g.}, a developer that is experienced outside the project
may already be known due to outside channels, and thus have an inflated likelihood of being @-mentioned to begin with.
We consider an outside-project attribute, developer's GitHub age (in days), in order to control for experience
outside the project which may lead to increased @-mentions when project contributions are relatively low. As GitHub age is readily observable through
the profile interface on GitHub (\emph{e.g.}, by viewing the contribution heatmap), 
we believe this to be a reasonably observable control.
Another outside-project control we considered was number
of public repositories contributed to by a developer, as this is readily observable; however, this was highly correlated with age, and was thus
dropped from the model.

\head{Modeling Future @-Mentions}
\label{sec:data_mention_model}
To answer our questions, we use count regression in a predictive model. This allows us to inspect the relationship
between our response (\emph{dependent variable}) and our explanatory variables (\emph{predictors} or \emph{covariates}, \emph{e.g.}, responsiveness) under the effects of various \emph{controls} (\emph{e.g.}, project size).

There are many forms of count regression; most popular are so-called Poisson, quasi-poisson, and negative binomial regression, all which
model a count response. In our work, we are interested in @-mentions as measured by number of incoming @-mention calls per person - a count.
In addition, as our data contain many zeros, we need a method that can accommodate;
the methods listed above all have moderate to severe problems with modeling zeros.
\emph{Zero inflated negative binomial regression} and \emph{hurdle regression} are two methods specifically designed to address
this challenge by explicitly modeling the existence of excess zeros~\cite{cameron2013regression}. It is common to fit both types
of models, along with a negative binomial regression, and compare model fits to decide which structure is most appropriate.
Standard analysis of model fit for these methods uses both Akaike's Information Criterion (AIC) and Vuong's test of non-nested
model fit to determine which model works best~\cite{vuong1989likelihood}.

We employ \emph{log} transformations to stabilize coefficient estimation and improve model fit, when appropriate~\cite{cohen2013applied}.
We remove non-control variables that introduce \emph{multicollinearity} measured by \emph{variance inflation factor (VIF)} $> 4$ (\emph{e.g.}, we do not use $\mathcal{ISS}_{\rho}$ due to high VIF),
as multicollinearity reduces
inferential ability; this is below the generally recommended maximum of $5$ to $10$~\cite{cohen2013applied}. Keeping control variables
with high VIF is acceptable, as collinearity affects standard error estimates; as control variables are not interpreted,
we do not much care if their standard error estimates are off~\cite{allisonMulticollinearity}.
We model on the person-project level, \emph{i.e.}, each observation is a person within a project.
We performed multiple hypothesis testing (p-value) correction by the Benjamini and Hochberg method~\cite{benjamini1995controlling}.
A squared age term is present in the zero model to account
for a quadratic shape in the residuals, along with its lower order term as is standard in regression~\cite{faraway2014linear}.

As noted in Section~\ref{sec:research_questions}, we explicitly model future @-mentions; our response variable
is the value $6$ months after our ``observed'' (\emph{i.e.}, covariate) data. As such, we build a \emph{predictive} model,
not a fully regressive model - \emph{i.e.}, one that is built on the entirety of available data. We note the difference is minor,
but worth reiterating.


\section{Results and Discussion}
\label{sec:results_and_discussion}

\noindent \textbf{Case Study: Project-Level Reasons for Call @-Mentions}
\label{sec:case_study}

We are interested in empirically measuring the reasons behind the @-mention. To  make sure our
theoretical underpinnings are reasonable,
we performed a random manual inspection
of $100$ call @-mentions from our data set, to qualitatively identify the primary reason behind the call.
 The counts of each category found in our manual inspection is shown in Table~\ref{tab:case_study_calls}.
In the case of \textbf{R}, we argue that reliance and/or trust in the mentionee is clear: the mentioner explicitly requests that the mentionee
performs some defined task; if the mentionee was deemed unreliable, the mentioner would be unlikely to trust them with an explicit task.

For \textbf{R-S}, the mentionee is not explicitly
called upon to perform some task. However, the mentioner seems to want the mentionee to respond (or perform a task), but does
not wish to explicitly tell the mentionee to act, likely out of politeness. Though the call to action is not explicit, we argue this
still represents mentionee reliability; like \textbf{R}, the mentioner wants the mentionee to perform an action, but does not explicitly state as much.

In the case of \textbf{I}, the call is meant to tag the @-mentionee in case they want to participate; not necessarily in order
to respond to the thread, or perform some action. However, the mentioner believes
that the mentionee may be interested in the issue at hand. This is similar to \textbf{R-S}, albeit slightly weaker,
as the mentioner may not have a particular task in mind for the mentionee. However, this still indicates that participation
 from the mentionee may be appreciated. 

In the case of \textbf{CA}, the mentioner is calling the mentionee in order to give credit, \emph{e.g.}, when the mentionee produced an important
patch that is relevant to the discussed issue. Though this is not a clear reliance on the mentionee in description, in practice we find it is often used
in a similar way to \textbf{I}; participation from the mentionee may be appreciated, but not necessary.

Across all $100$ manually inspected cases, we found only $3$ cases in which the call @-mention does not fall into the aforementioned categories ($3\%$);
one appears to be a misuse of the @-mention;
the other two are due to users changing their GitHub display name after the @-mention is seen, thus throwing off our detection of the @-mention as
a call rather than a reply. Thus, we argue that the call @-mention is consistently representative of reliance on the mentionee.

\begin{table}
\centering
\caption{Call @-mention categories, samples, and case study.}
\label{tab:case_study_calls}
\resizebox{0.85\columnwidth}{!}{%
\begin{tabular}{|c|c|c|}

\hline
Category & Example & Count \\
\hline R & 
  \begin{tabular}{@{}p{65mm}@{}}
    \emph{Project: hashicorp/terraform; Issue: $7886$} \\
    ``@phinze - can we please have someone take a look 
      at this PR now that tests and docs are complete?''
  \end{tabular} 
              & $39$ \\
\hline R-S & 
  \begin{tabular}{@{}p{65mm}@{}}
    \emph{Project: dotnet/roslyn; Issue: $18363$} \\
    ``''... I don't know if the test flavor recognizes this capability. 
    @codito @sbaid would know.''
  \end{tabular} 
              & $17$ \\
\hline I & 
  \begin{tabular}{@{}p{65mm}@{}}
    \emph{Project: dotnet/corefx; Issue: $8673$} \\
    ``/cc @DavidGoll @karelz 
      My current understanding (based on 
      WinHTTP's response) is ...''
  \end{tabular} 
              & $33$ \\
\hline CA & 
  \begin{tabular}{@{}p{65mm}@{}}
    \emph{Project: avajs/ava; Issue: $1400$} \\
    ``... There is already a PR for this though, thanks 
      to @tdeschryver ...''
  \end{tabular} 
              & $8$ \\
\hline \hline Misuse or Misclassification & 
  \begin{tabular}{@{}p{65mm}@{}}
    \emph{Project: celery/celery; Issue: $817$} \\
    ``We are also using them in production @veezio for
      quite some time, works fine.'' \\
    \emph{Author's note: @veezio is a company GitHub account.}
  \end{tabular} 
              & $3$ \\
\hline
\end{tabular}
}
\end{table}


\subsection{Future @-Mention Models}
\begin{figure}
\includegraphics[width=0.85\columnwidth]{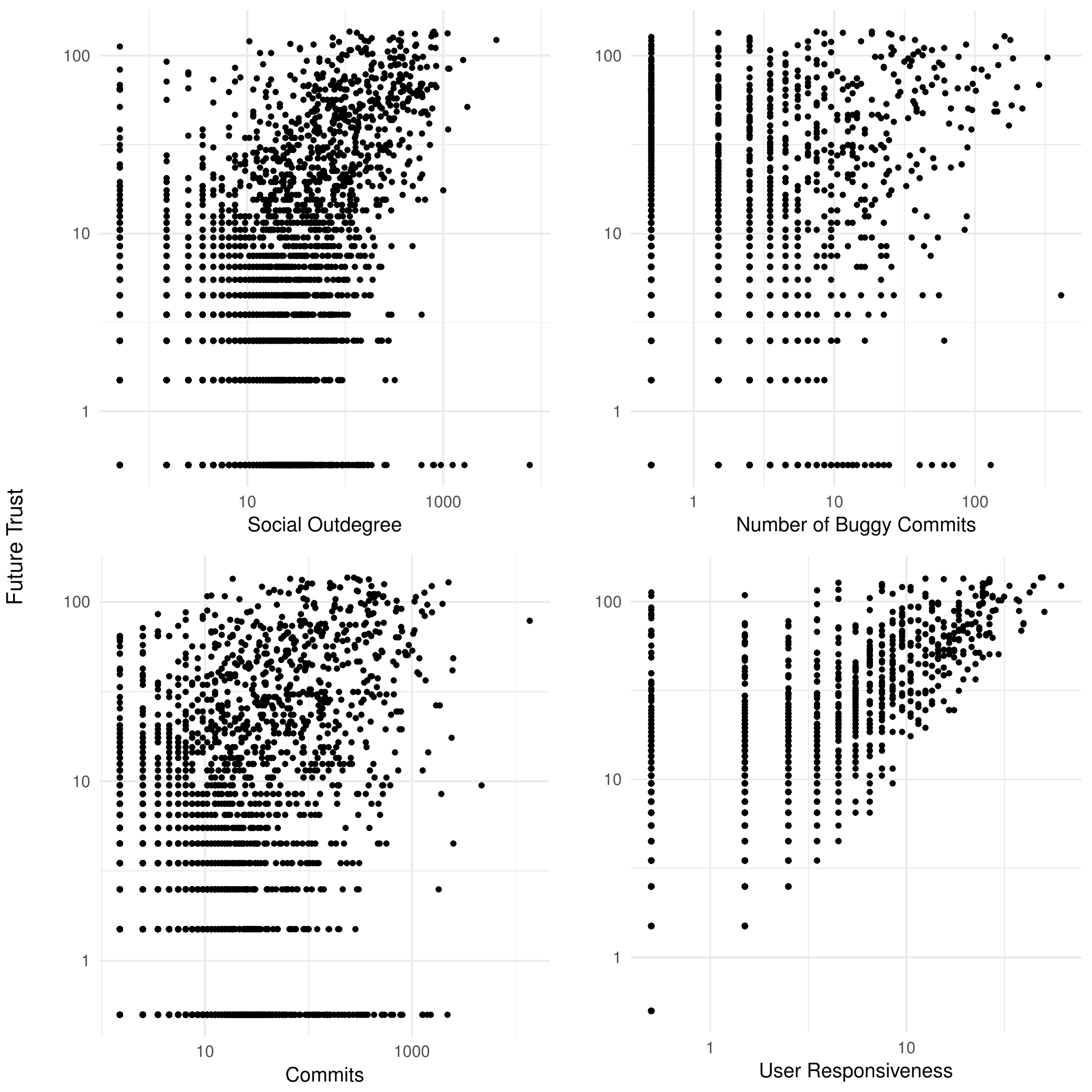}
\caption{Future @-mentions vs. selected attributes of visibility, expertise, productivity, and responsiveness. Axes log scaled.}
\label{fig:exploratory_scatter_plots}
\end{figure}

Fig.~\ref{fig:exploratory_scatter_plots} shows a selection of variables from our categories of interest and their paired
relationship with future @-mentions. For all variables, we see a strong positive relationship with @-mentions; the largest
correlation sits with developer responsiveness ($78.90\%$).

Though paired scatter plots provide initial insight to affecters of potential power, we must model them
in the presence of other variables, along with controls, to properly answer our  questions.

\input{model_new.tex}

\begin{figure}
\includegraphics[width=.7\columnwidth]{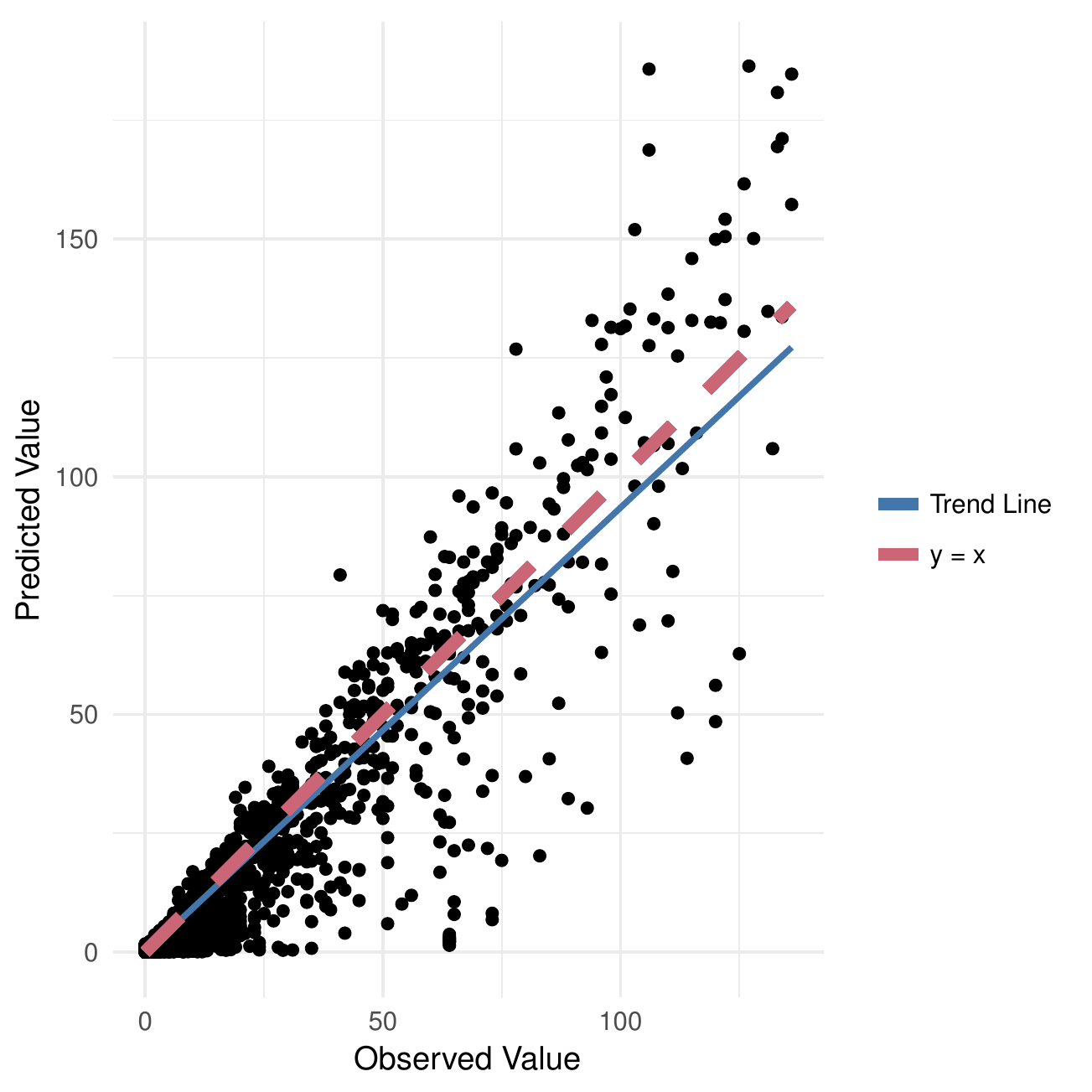}
\caption{Predicted vs. observed values.}
\label{fig:model_pred_obs}
\end{figure}

\rqone

Table~\ref{tab:model_future_trust} shows our model of future @-mentions, with affecters of interest grouped and separated from one another.
Our analysis points to a zero hurdle model as providing the best fit, which separately models the process of attaining one's first call
(``zero'' model, logistic regression), and the process of attaining beyond one call (``count'' model, Poisson regression).
Fig.~\ref{fig:model_pred_obs} depicts predicted and observed values along with a $y=x$ and trend line.
The mean average and mean squared error are $0.910$ and $15.769$, respectively, indicating a good average model fit.

\head{Visbility} We see that \OSS{\rho} and social outdegree are positive for both the count and zero components of our model.
This suggests that a higher social focus (in replying to others) and larger overall social outdegree associates with 
being @-mentioned in the future - be it in the transition from zero to greater than zero @-mentions, or in increasing @-mentions.
However, we see a negative coefficient for \ISS{\kappa},
suggesting that when others focus their calls on the observed individual, the observed's @-mention count decreases\footnote{\ISS{\kappa} is not
used for the zero component; it is undefined when call mentions are $0$.}. This negative coefficient is not unexpected; \ISS{\kappa} is derived
from the Kullback-Leibler divergence, and when there are many cells (\emph{i.e.}, others that can be called), it is expected that a higher focus
is correlated with a lower raw value. \emph{E.g.,} consider the case where  $10$ individuals can call on developer $A$. If each calls $A$
once, the raw value for calls is $10$ and \ISS{\kappa} is low; if only one developer calls $A$, the raw value is $1$ but \ISS{\kappa} is high. In support
of this intuition,  Posnett \emph{et al.}~\cite{posnett2013dual} found that a
higher value of $\mathcal{DAF}$ associates with a lower raw cell count. 

In sum, having a larger social presence (\OSS{\rho}, social outdegree) may associate with one's future @-mention count. These values are much easier to
increase for an individual than \ISS{\kappa}, as \ISS{\kappa} is a function of indegree,
and thus less in the individual's control.

\head{Expertise} The number of likely buggy commits a developer makes has a negative coefficient for the count component, suggesting 
that a larger number of likely buggy commits associates with a decrease in @-mentions.
This is as expected: a higher expertise should lead to more future @-mentions.
However, we see a positive coefficient for the zero component. This is puzzling at first, but may be explained thusly: it is known that
contributions are extremely important in order to gain technical trust in OSS~\cite{gharehyazie2015developer}, 
supported also by the large coefficient for commits in the 
zero component ($0.453$). As the number of likely buggy commits is correlated with the number of overall commits by a developer, this positive 
coefficient
indicates that contributing at all, regardless if one's contribution is buggy, is important in getting the first call mention, and thus
the first @-mention.

\head{Productivity} In both the zero and count components, we see a positive coefficient for commits, indicating that increased productivity is
associated with higher @-mentions. The zero model coefficient is very high.
This is in support of productivity being important in receiving the first @-mention.

\head{Responsiveness} Interestingly and contrary to our hypothesis, for the count component,
we see an insignificant coefficient. Responsiveness is not considered in the zero component as one must be called in order to reply, which means
responsiveness is undefined for those with an @-mention count of $0$.

\vspace{-1.5mm}
\ra{1}{We see a positive effect of visibility measured by \OSS{\rho}, and a negative effect for \ISS{\kappa}. Also, more likely buggy commits (a measure of negative expertise) 
is associated with lower @-mentions when one has already been @-mentioned, and higher @-mentions if one has not yet been @-mentioned,
 possibly explained by the idea that any productivity associates with a first @-mention.
We see positive effects for productivity, and no significant effect of responsiveness.}

\vspace{1mm}
\noindent \textbf{Case Study: Attributes of Interest and Model Fit.}

To further examine \textbf{RQ 1} and provide concrete reasoning behind our model's fit, we performed case studies. Specifically,
we looked at those with high observed future @mentions but low model predictions, and those who transition from zero to nonzero @-mentions.

\noindent \textbf{Sub-Case Study: High Observed @-Mentions, Low Predicted @-Mentions}.

For this study, we manually examined those with less than $50$ and greater than $15$ 
observed future @-mentions, nonzero observed @-mentions, and a predicted
@-mention count of less than or equal to $1$; \emph{i.e.,} those along the bottom of the x-axis of Fig.~\ref{fig:model_pred_obs}.
In this region, all individuals have never explicitly replied to another developer (\emph{i.e.}, \OSS{\rho} and social outdegree are both $0$), a
low number of commits ($1$ to $9$); as these coefficients are positive in our model, these individuals should be pushed to higher counts. However, all
developers in this region also have relatively high \ISS{\kappa} ($0.1$ to $1.0$), and have experience in other projects (indicated by a large developer age).
As both \ISS{\kappa} and developer age have a relatively large negative influence in our model, this
explains why our predicted future @-mentions are low from a statistical standpoint.

To dig deeper, we consider the case of a particular developer in this region: developer \emph{arthurevans}, for project \emph{google/WebFundamentals}.
In issue $\#4928$ of the project,
a discussion about
PRPL patterns\footnote{\url{https://developers.google.com/web/fundamentals/performance/prpl-pattern/}}, the poster says:
``\emph{I'll defer to the grand master of all things PRPL, @arthurevans for what the final IA for this section might look like}''. 
Although \emph{arthurevans} has low observed activity in the project itself (\emph{e.g.}, low social outdegree and low commit count),
this indicates that the poster greatly values \emph{arthurevans}'s input. The story is similar for the others in this region
\footnote{We could not perform this in-depth study for  discussions not in English.};
the issue poster values the opinion of the called-in person, indicating a level of outside-project expertise. 

In summary, it appears this region consists of those who are actually expert, but this expertise is not reflected by their in-project contributions.
Although we attempt to capture
outside expertise through a developer's overall GitHub age, we were unable to include other metrics of outside expertise (\emph{e.g.} number of public
repositories contributed to) due to high multicollinearity. Orthogonal metrics of outside expertise may exist that can better fit these
individuals.

\noindent \textbf{Sub-Case Study: Transitioning From Zero @-Mentions}.

For this study, we took a random sample of $10$ individuals (out of $235$) who had zero observed @-mentions,
but transitioned to nonzero @-mentions in the next $6$ months \emph{i.e.}, our future period.
In this region, we observe a combination of factors: project age and newcomers who wish to participate more.
Some projects are relatively new or newly popular, which means that although they are rapidly gaining popularity on GitHub,
their issue production rate hasn't yet caught up. Though all individuals have
contributed to the project, there has not been a chance for
@-mentions to be observed; those transitioning from zero @-mentions to nonzero @-mentions would likely have nonzero @-mentions had the observation
time split been later in the project.

Perhaps more interesting, we see some new individuals that have recently contributed commits and seem genuinely interested in participating
more. For example, in pull request $\#2587$ of the project \emph{prometheus/prometheus}, we see the first call to developer \emph{mattbostock}, causing
a transition from zero to nonzero @-mentions. Prior to this, we see that \emph{mattbostock} had been contributing to issue discussions
(\emph{e.g.} issues $\#1983$ and $\#10$), bringing up problems and providing potential solutions. Thus, due to signaling interest and
participating in discussions (visibility), providing commits (productivity), and having no bugs in these commits (expertise), we see
that the fruits of their labor have resulted in an @-mention.

\rqtwo

\begin{figure*}
\centering
\begin{subfigure}{.85\columnwidth}
\includegraphics[width=\columnwidth]{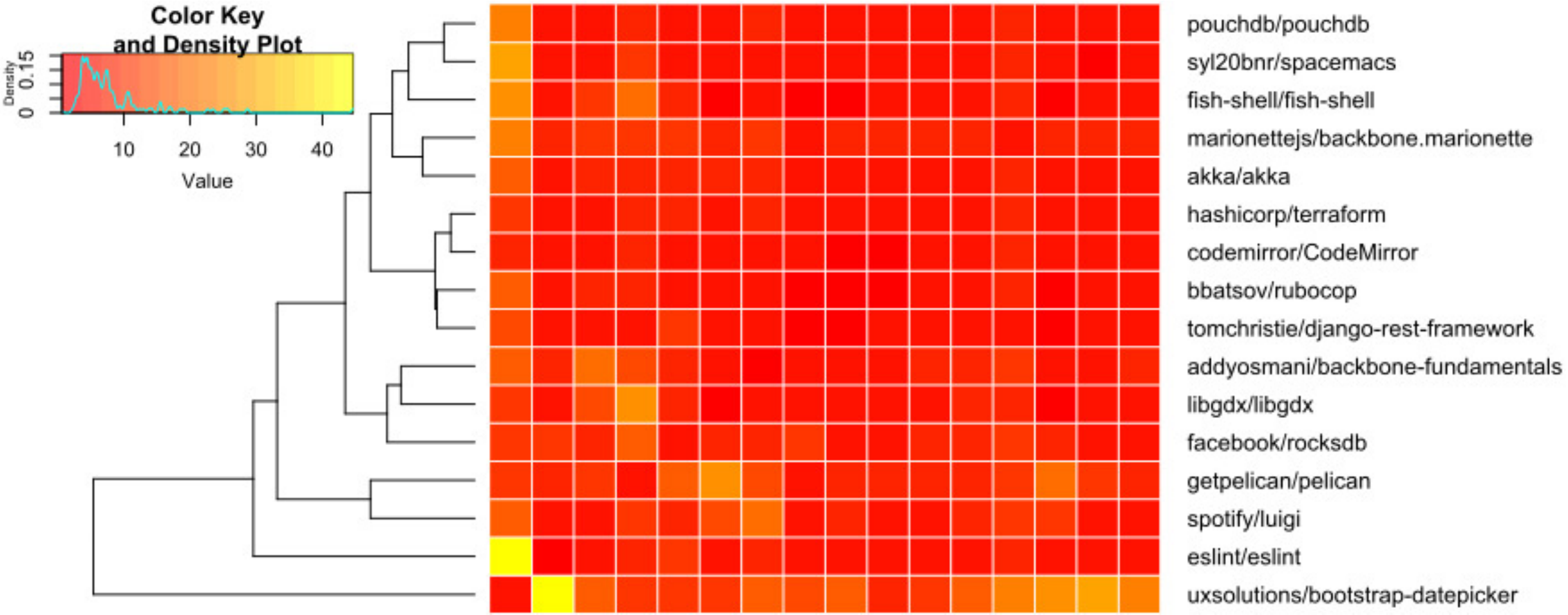}
\end{subfigure}
\qquad 
\begin{subfigure}{.85\columnwidth}
\includegraphics[width=\columnwidth]{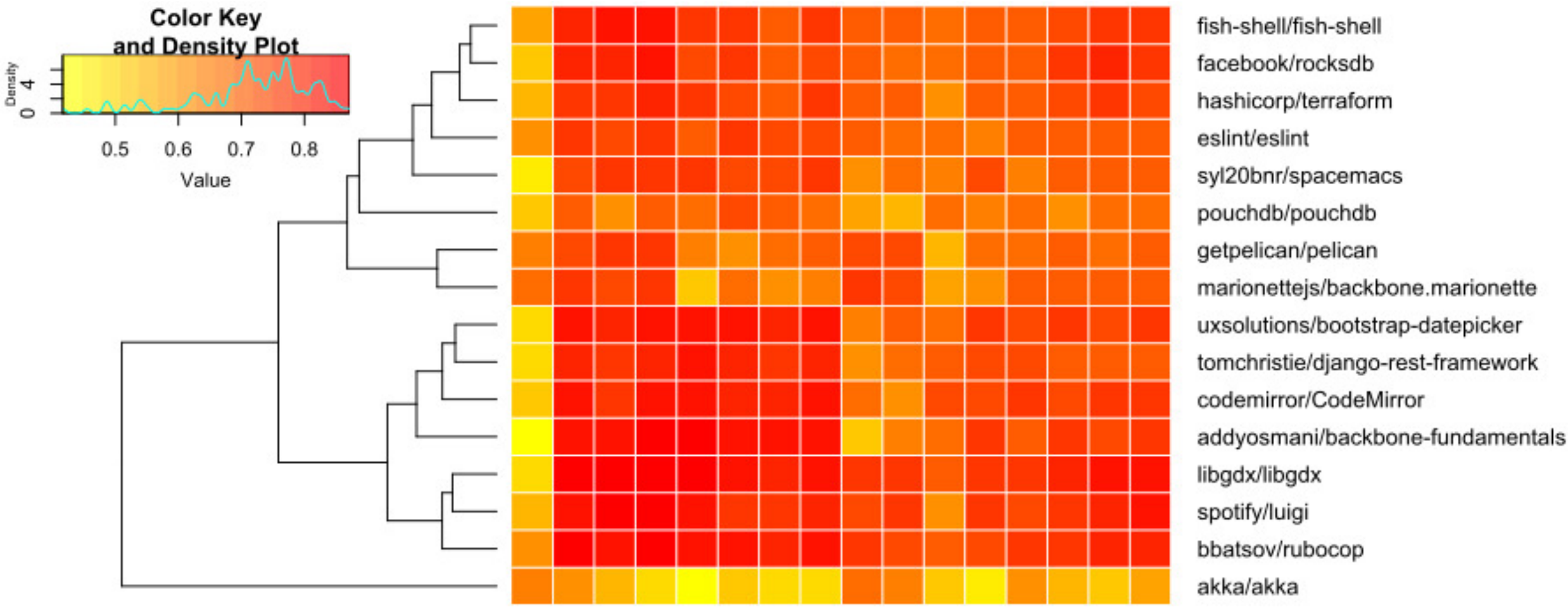}
\end{subfigure}
\caption{Cross-project predictive power heatmap for each project-specific model, count (left) and zero components.}
\label{fig:heatmap_predictability}
\end{figure*}

To answer this question, we require project-specific models of @-mentions. Due to the sparseness of data, adding a factor
to the existing model in Table~\ref{tab:model_future_trust} causes estimation to diverge. Thus, to avoid divergence, we fit simplified models
with selected attributes of visibility (\OSS{\rho}, \ISS{\kappa}, social outdegree), expertise (likely buggy commits), productivity
(commits), responsiveness, and developer's outside project experience (GitHub age). A subset is required due to the smaller
number of observations per project; too many variables for too little data can cause issues as, \emph{e.g.},
small multicollinearity can cause big issues for small data. Thus, we select only a few representative variables from each
of our groups of interest. For consistency, we explicitly fit separate models
for the transition from zero to nonzero (zero component) and for nonzero count (count component), as is done
implicitly by the hurdle model.

Fig.~\ref{fig:heatmap_predictability} contains symmetric heatmaps of predictability for our project-specific models (count and zero,
respectively). To measure predictability of the count component, we use the average of \emph{mean absolute error} (MAE) between each
pair of models. For projects $i$ and $j$, with data $d_i$ and $d_j$, and models $y_i$ and $y_j$, we compute predicted values
$\hat{y}_i = y_i(d_j)$ and $\hat{y}_j = y_j(d_i)$; \emph{i.e.} we predict using one model's fit and the other model's data,
thus providing a measure of cross-project model fit. We then compute
the average MAE between the two fits \emph{i.e.}, $\frac{\hat{y}_i + \hat{y}_j}{2}$, and plot this value in each heatmap cell.
For the zero component, we analogously compute fit by calculating the average \emph{area under the receiver operating characteristic curve} (AUC)
between two projects \emph{i.e.} $\frac{AUC(\hat{y}_i) + AUC(\hat{y}_j)}{2}$. For MAE, a lower value is better; for AUC, a higher value.
We then plot a dendrogram, showing clusters of projects based on predictive ability.

For both the count and zero components, we generally see good fit across projects (lower average MAE, higher average AUC), with some outliers.
For the count case, we see that \emph{uxsolutions/bootstrap-datepicker} is an anomaly in having poor fit for many projects,
being grouped in its own cluster. Otherwise, there are no immediately clear clustering relationships between projects, other than that the
mean MAE is generally below $10$, as noted in the density plot.

\begin{figure*}
\centering
\begin{subfigure}{.8\columnwidth}
\includegraphics[width=\columnwidth]{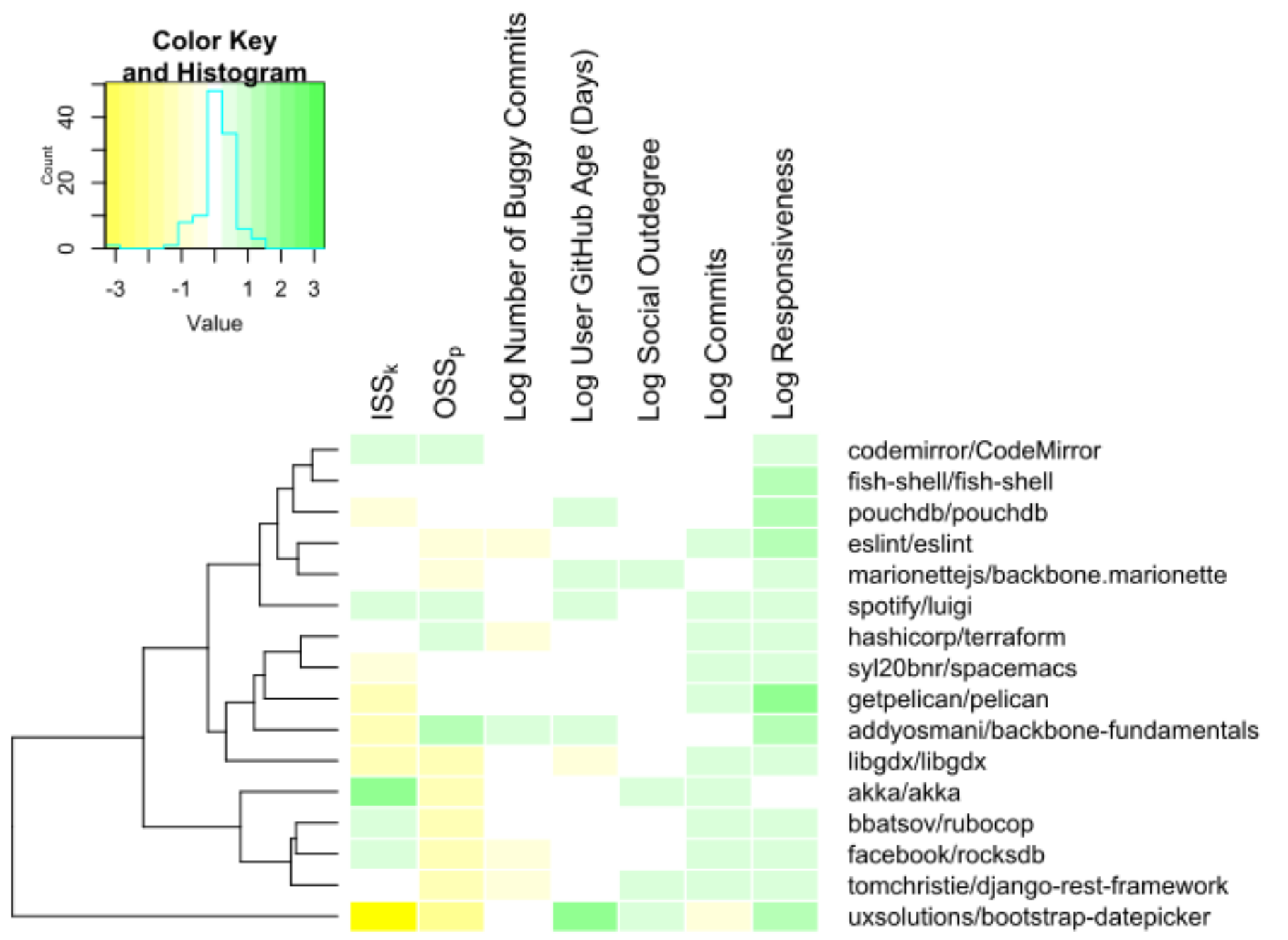}
\end{subfigure}
\qquad
\begin{subfigure}{.8\columnwidth}
\includegraphics[width=\columnwidth]{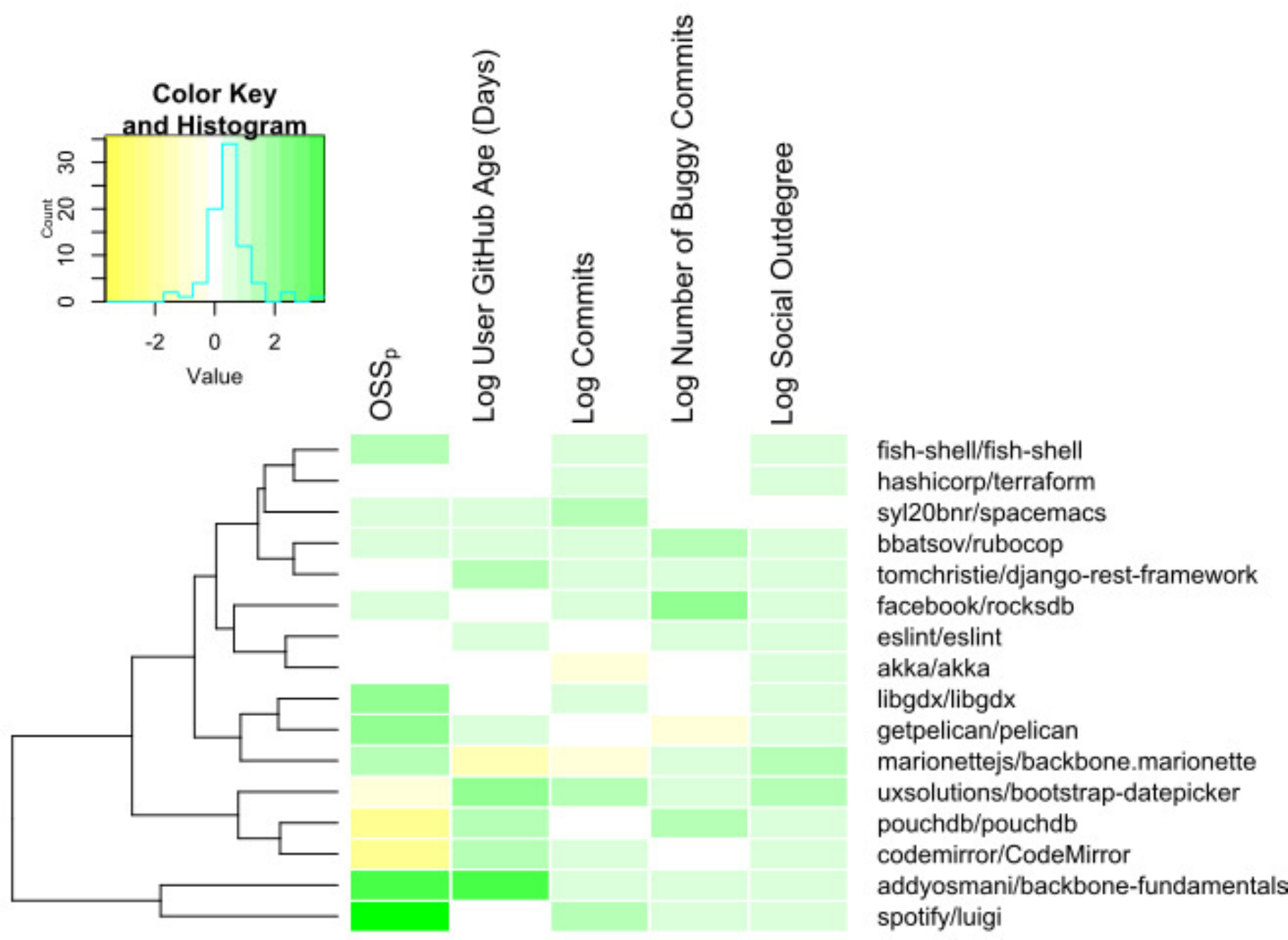}
\end{subfigure}
\caption{Heatmap of coefficients for project-specific models, (a) count and (b) zero components.}
\label{fig:heatmap_coefficients}
\end{figure*}

For the zero case, we also see one clear outlier: \emph{akka/akka}. In general, cross-project fits for this project are relatively poor
compared to the majority. The reason for this may be due to the difference in importance for our affecters of interest as compared to other projects.
Fig.~\ref{fig:heatmap_coefficients} shows our fitted coefficients for each project model. For the zero component,
though \emph{akka/akka} does not lie on its
own according to hierarchical clustering, we see that its coefficients are very different from other projects, with a negative coefficient for
commits and almost zero coefficients for all other variables (except social outdegree). This explains the poor cross-project fit; in this project,
a higher number of commits associates with a lower predicted @-mention count, while in the majority of other projects this
coefficient is positive (or nearly zero).

In summary, we do see a general trend of good fit for both the count component and, to a lesser extent, the zero component.

\vspace{-1.8mm}
\ra{2}{We see that the count component of each project-specific model has overall good fit when predicting purely cross-project. We see a similar
trend for the zero component, though to a lesser extent on average.}
\vspace{1.5mm}

\rqthree

Fig.~\ref{fig:heatmap_coefficients} contains heatmaps
of coefficients for the count and zero components of our project-specific models. When looking at each column, we see
some coefficients that are almost uniformly the same, \emph{e.g.}, responsiveness for both components, commits for the count component,
and likely buggy commits for the zero component. However, we do see differences, \emph{e.g.}, \OSS{\rho} in both model
components is negative for some and positive for other rows.

The fact that there are differences per column (\emph{i.e.}, per coefficient) for most coefficients lends credence to the idea
that there are project-specific @-mention culture differences on a per attribute basis. However, there are things that
don't change across projects, \emph{e.g.}, the importance of commits in gaining more @-mentions.
In addition, the generally high cross-project predictive power shown in Fig.~\ref{fig:heatmap_predictability} suggests that project-specific culture differences may not matter too much.
To identify some concrete reasoning behind these particular differences in variable importance, we turn to another case study.

\noindent \textbf{Case Study: Project-Level Differences.}

Reflecting on Fig.~\ref{fig:heatmap_coefficients}, here we ask: why are some coefficients positive
for a number of projects, and negative for others? 

As \OSS{\rho} seems to exhibit this behavior in both our count and zero models,
and significantly so for our global model, we choose it for our study.
For the zero model, we see a negative coefficient for projects \emph{uxsolutions/bootstrap-datepicker, pouchdb/pouchdb}, 
and \emph{codemirror/CodeMirror}; indicating  higher specialization in one's replies associates with lower
future @-mentions in the projects.

One explanation for this phenomenon could be due to a larger inner
circle
as compared to other projects;
\emph{i.e.}, to gain @-mentions one must become visible to more people. For both \emph{uxsolutions/bootstrap-datepicker} and
\emph{pouchdb/pouchdb}, this seems to be the case. When looking at the distribution of commits across contributors, in both projects
the original top committer has largely reduced their commit rate, while in the mean time the second largest committer has picked up the pace.
In addition, the distribution of commits seems to be comparatively more uniform across contributors, indicating a larger inner circle.
For \emph{codemirror/CodeMirror}, the distribution of commits is highly concentrated in the top committer; however, when viewing issues,
we see that multiple others contribute to review and discussion. This likewise indicates a larger inner circle that one must be visible to.
For the count model, the story seems to be the same for projects with a negative coefficient; there is either a more uniform distribution
of commits across the top contributors, or a larger number of individuals participating in issue discussions, indicating a larger
inner circle.

For projects with positive coefficients, we see a different behavior. In pull requests, it appears the top project members
are more open to calling on others to provide input. \emph{E.g.,} for project \emph{spotify/luigi} pull request $\#2186$, a top contributor
asks the original poster to run \emph{git blame} on the modified code to see who originally posted it, admitting a lack of expertise
about the associated module; we see similar behavior for pull request $\#2185$. For project \emph{addyosmani/backbone-fundamentals} issue $\#517$,
we see the project owner calls on another contributor for their input, stating ``\emph{[I] would love to suggest your project to devs ...}''.
Recall that a positive coefficient for \OSS{\rho} indicates a specialization in reply behavior, \emph{i.e.}, more focus in one's social behavior.
As the top contributors for these projects seem to be the ones calling on others, it appears one may specialize their social behavior towards
the top contributors to get noticed; hence, more social specialization may associate with higher future @-mentions. 

\vspace{-1.8ex}
\ra{3}{We see slight indications of project-specific @-mention culture.
The high cross-project performance suggests that these differences may not
matter much for predictivity.}
\vspace{-1.5ex}





\section{Conclusion and Threats}
\label{sec:conclusion}
We performed a  quantitative study of @-mentions in GitHub, as captured in calls to people in discussions.
We supplemented those with case studies on samples of discussions, to help triangulate our findings.
The well-fitting, reasonable models, suggest that our formulation of @-mentions is explained well by the data. 

Some of our results were less obvious than others, \emph{e.g.}, the insignificant effect of responsiveness and the positive effect of  commits that get fixed on the initial @-mention.
From a security perspective, it may follow then, that trusting new people with the projects code is associated with more code that needs fixing, perhaps via changes that they may introduce, which is certainly a concern. Based on these results, increased efforts towards training new people to the specifics of the project's code, \emph{e.g.}, in creating a portal for newcomers~\cite{steinmacher2016overcoming},
can be appropriate.


The idea that projects in an ecosystem have similar models of what it means to be worthy of an @-mention is appealing.
We find that the good cross-project predictive power cannot be simply distilled down to productivity in our models, thus 
adding evidence toward the multidimensional nature of @-mentions.
It is also very reasonable that there would be cliques of projects in which the sense of who to @-mention is even more uniform than across the whole ecosystem, and our findings underscore that.
Obvious open questions here are: how do notions of @-mentions get in sync? And,  to borrow from ecology, does the robustness of the @-mention models across GitHub convey any fitness benefit in the ecosystem?
We can see a plausible mechanism that would offer an answer to the first: projects share people and people cross-pollinate the @-mentioning behavior across projects in which they participate. We leave the validation of this, and other models, to future work. 
The @-mention model robustness, likewise, implies some preference for success, be it by design or an emerging one, across the ecosystem.
This can be a function of people's mobility in the ecosystem and their preference for and vigilance to participate in popular projects; we leave the answers for future work.

\head{Threats to Validity}
There were challenges involved in all aspects of the work, largely due to the loaded reasoning behind @-mentions. Being @-mentioned is not just a result
of technical prowess; @-mentioning is also a social phenomenon.
Many potential issues were anticipated and carefully addressed.
Once we settled on the idea of using call @-mentions, we were able to connect
our outcome with background theory on the multidimensionality of @-mentions.
To define @-mentions precisely, we necessarily had to narrow our definition specifically to call mentions in issue discussions.

Our case studies would benefit from larger amount of data. The case study sizes were
due to the regions of interest; our regions were small, and thus our case studies were relatively small.

Our work is supported by prior qualitative research into @-mention usage. Still, we acknowledge that our study would likely benefit from further qualitative studies, \emph{e.g.}, a survey of developers on their use of the @-mention.


\end{document}

%% file: abstract.tex

Open Source Software (OSS) project success relies on crowd contributions.
When an issue arises in pull-request based systems, @-mentions are used to call on people to task;
previous studies have shown that @-mentions in discussions are associated with faster issue resolution.
In most projects there may be many developers who could technically handle a variety of tasks. 
But  OSS supports dynamic teams distributed across a wide variety of social and geographic backgrounds, as well as levels of involvement.
It is, then, important to know whom to call on, \emph{i.e.}, who can be relied or trusted with important task-related duties, and why. 

In this paper, we sought to understand which observable socio-technical attributes of developers can be used to build good models of them being future @-mentioned in GitHub issues and pull request discussions.
We built overall and project-specific predictive models of future @-mentions, in order to capture the determinants  of @-mentions in each of two hundred GitHub projects, and to understand if and how those determinants differ between projects. 
We found that visibility, expertise, and productivity are associated with an increase in @-mentions, while responsiveness is not, in the presence of a number of control variables. Also, we find that though  project-specific differences exist, 
the overall model can be used for cross-project prediction, indicating its GitHub-wide utility.

%% file: model_new.tex
\begin{table}[!htbp] \centering 
  \caption{Future @-mention model; p-values corrected by BH method. User subscripts omitted; they refer to the developer under observation within the model} 
  \label{tab:model_future_trust} 
\resizebox{\columnwidth}{!}{%
\setlength{\tabcolsep}{1ex}
\begin{tabular}{@{\extracolsep{5pt}}lcccc} 
\\[-1.8ex]\hline 
\hline \\[-1.8ex] 
 & \multicolumn{4}{c}{\textit{Dependent variable:}} \\ 
\cline{3-4} 
\\[-1.8ex] & \multicolumn{4}{c}{Future @-mentions (6 months later)} \\ 
\\[-1.8ex] & Count & (Std. Err.) & Zero & (Std. Err.)\\ 
\hline \\[-1.8ex] 
\textit{Visibility} & & \\ \cline{1-1} \\ [-1.8ex]
 $\mathcal{OSS_{\rho}}$ & 0.103$^{*}$ & (0.045) & 0.351$^{***}$  & (0.100) \\ 
  
 $\mathcal{OSS_{\kappa}}$ & $-$0.046 & (0.040) & 0.508$^{***}$  & (0.099) \\ 
  
 $\mathcal{ISS_{\kappa}}$ & $-$0.283$^{***}$ & (0.047) &  & \\ 
 Log Social Outdegree & 0.058$^{***}$ & (0.008)& 0.433$^{***}$  & (0.022) \\
 \\[-1ex]
\textit{Expertise} & & \\ \cline{1-1} \\ [-1.8ex]
 Log Number of Buggy Commits & $-$0.065$^{***}$ & (0.010) & 0.187$^{***}$  & (0.043) \\ 
  
 $\mathcal{DAF}$ & $-$0.040 & (0.042) & $-$0.134 &  (0.101) \\ 
  
 Top Committer or Project Owner & 0.055 & (0.044) & 0.691  & (0.534) \\ 
  \\[-1ex]
\textit{Productivity} & & \\ \cline{1-1} \\ [-1.8ex]
 Log Commits & 0.086$^{***}$ & (0.008) & 0.453$^{***}$  & (0.025) \\ 
  \\[-1ex]
\textit{Responsiveness} & & \\ \cline{1-1} \\ [-1.8ex]
 Log User Responsiveness & $-$0.003 & (0.012) &  & \\ 
  \\[-1ex]
\textit{Controls} & & \\ \cline{1-1} \\ [-1.8ex]
 Committer Only & 0.141$^{***}$  & (0.039) & $-$1.584$^{***}$  & (0.060) \\ 
  
 Log Total Posts in Project & 0.021$^{*}$ & (0.010) & 0.151$^{***}$  & (0.021) \\ 
  
 Log Observed @-Mention Value & 0.980$^{***}$ & (0.011) & & \\ 
  
 User GitHub Age (Days) & $-$0.137$^{***}$ & (0.020) & $-$1.470$^{***}$  & (0.430) \\ 
  
 User GitHub Age (Days) Squared & & & 0.116$^{***}$ & (0.031) \\ 
  
 Intercept & 0.637$^{***}$ & (0.180) & 1.684  & (1.511) \\ 
  
\hline \\[-1.8ex] 
Observations & 17,171 & & \\ 
Mean Absolute Error & 0.910 & & & \\
Mean Squared Error & 15.769 & & &\\ 
\hline 
\hline \\[-1.8ex] 
 \multicolumn{5}{r}{$^{\dagger}$p$<$0.1; $^{*}$p$<$0.05; $^{**}$p$<$0.01; $^{***}$p$<$0.001} \\ 
\end{tabular} 
}
\end{table} 